\documentclass[12pt]{article}
\newcommand{\layout}{manuscript}

\usepackage[latin1]{inputenc}
\usepackage{amsmath}
\usepackage{amsfonts}
\usepackage{amssymb}

\usepackage{amsmath}
\usepackage[sort&compress]{natbib}
\usepackage{ifthen}

  \usepackage{graphicx}
  \usepackage{epstopdf}
   \newcommand{\epspdf}{pdf}

\renewcommand{\rho}{\varrho}

\renewcommand{\eqref}[1]{(\ref{#1})}

\newcommand{\figref}[1]{Fig.~\ref{#1}}
\newcommand{\Figref}[1]{Figure~\ref{#1}}

\DeclareMathOperator{\gra}{\mathrm{grad}}

\renewcommand{\div}{\mathrm{div}\,}
\DeclareMathOperator{\lap}{\triangle}

\newcommand{\vek}[1]{\boldsymbol{#1}}

\newcommand{\diffI}[2]{\frac{\partial #1}{\partial #2}}
\newcommand{\diffII}[2]{\frac{\partial^{2} #1}{\partial #2^{2}}}

\newcommand{\Kl}{\left ( }
\newcommand{\kl}{\right )}
\newcommand{\Ekl}{\left [}
\newcommand{\ekl}{\right]}
\newcommand{\Gkl}{\left\lbrace }
\newcommand{\gkl}{\right\rbrace}

\ifthenelse{\equal{\epspdf}{pdf}}{
  \ifthenelse{\equal{\layout}{PNAS}}{
    \setlength{\voffset}{-1.3in}
    \setlength{\pdfpageheight}{29.8cm}
    }{
    \setlength{\voffset}{-1.5in}
    }
  }{
  \setlength{\voffset}{-1in}
  }
\ifthenelse{\equal{\layout}{PNAS}}{ \setlength{\oddsidemargin}{1cm} 
}{}
\ifthenelse{\equal{\layout}{manuscript}}{ 
\setlength{\oddsidemargin}{3cm} }{}

\ifthenelse{\equal{\layout}{PNAS}}{
  \setlength{\headheight}{0cm}
  \setlength{\topmargin}{0cm}
  \setlength{\textheight}{27cm} 
  }{
  \setlength{\textheight}{24cm}
  \setlength{\headsep}{1.5cm}
  \setlength{\topmargin}{1cm}
  }

\ifthenelse{\equal{\layout}{manuscript}}{ 
\setlength{\textwidth}{15cm} }{}
\ifthenelse{\equal{\layout}{PNAS}}{ \setlength{\textwidth}{19cm} }{}

\ifthenelse{\equal{\layout}{PNAS}}{
  
}{
  
}

\catcode`\@=11

  \def\fnum@figure{{\bf \figurename~\thefigure}}

  \def\fnum@table{{\bf \tablename~\thetable}}

\catcode`\@=12

\begin{document}

\begin{center}

  \ifthenelse{\equal{\layout}{manuscript}}{}{\today}

  \ifthenelse{\equal{\layout}{manuscript}}{ \vspace{1cm} }{}

  {\Large \bf Exact Solution for the Stokes Problem 

    of an Infinite Cylinder in a Fluid 

    with Harmonic Boundary Conditions at Infinity

  }

  \ifthenelse{\equal{\layout}{manuscript}}{ \vspace{1cm} }{}

  Andreas N. Vollmayr$^*$, Jan-Moritz P. Franosch$^*$, and J. Leo van 
Hemmen$^*$

  \ifthenelse{\equal{\layout}{manuscript}}{ \vspace{1cm} }{}

  $^*$ Physik Department T35, TU M\"{u}nchen, \\
  85747 Garching bei M\"{u}nchen,
  Germany \\
  Correspondence and request for materials
  should be addressed to \\ J. Leo van Hemmen
  (e-mail: LvH@tum.de).

  \ifthenelse{\equal{\layout}{manuscript}}{ \vspace{1cm} }{}



\end{center}

\section*{Abstract}

{\bf We present an exact solution for the time-dependent Stokes
  problem of an infinite cylinder of radius $r=a$ in a fluid with
  harmonic boundary conditions at infinity. This is a 3-dimensional
  problem but, because of translational invariance along the axis of
  the cylinder it effectively reduces to a 2-dimensional one. The
  Stokes problem being a linear reduction of the full Navier-Stokes
  equations, we show how to satisfy the no-slip boundary condition at
  the cylinder surface and the harmonic boundary condition at
  infinity, exhibit the full velocity field for radius $r > a$, and
  discuss the nature of the solutions for the specific case of air at
  sea level.}

\thispagestyle{empty}

\ifthenelse{\equal{\layout}{manuscript}}{\clearpage}{}

\section{Introduction: Navier-Stokes Equations}

We denote a position in space by $\vek{x}$ and time by $t$. We
describe an arbitrary fluid flow by its local velocity
$\vek{v}(\vek{x},t)$, density $\rho(\vek{x},t)$, and pressure
$p(\vek{x},t)$ and denote the total
derivative by
\begin{flalign}
 \frac{D \rho(\vek{x},t)}{Dt} := \diffI{\rho(\vek{x},t)}{t} + 
 \sum_{i=1}^3 v_i(\vek{x},t) \diffI{\rho(\vek{x},t)}{x_i} \,.
\end{flalign}
For the sake of clarity, we will omit the position and time dependency
and simply write $\rho = \rho(\vek{x},t)$. In an incompressible fluid
of dynamic viscosity $\nu_0=\mu/\rho$ and with volume force $\vek{F}$
the Navier-Stokes equations are given by \citep{Panton2005a}
\begin{flalign}
  \frac{D \vek{v}}{D t} = \diffI{\vek{v}}{t} + \Kl \vek{v} \cdot 
\nabla \kl
       \vek{v} = - \frac{1}{\rho_{0}} \gra p + \nu_{0} \lap
  \vek{v} + \frac{1}{\rho_{0}}\vek{F} \label{ImpulsInkomp}
\end{flalign}
while incompressibility is explicitly taken care of by
\begin{flalign}
  \div \vek{v}&=0 \,. \label{KontiInkomp}
\end{flalign}
Finally, we will use the no-slip boundary condition, meaning that the
velocity vanishes, i.e., $\vek{v}=\vek{0}$, at the surface of an
infinite cylinder whose axis we take parallel to the $z$-axis. The
cylinder radius equals $r=a$.

\begin{table}[hbt]
  \centering
  \begin{tabular}{crl}\\ \hline
    constant     & value & unit \\ \hline \hline
    $\nu_0$      & $15.11 \cdot 10^{-6}$ & $\mathrm{m}^2/\mathrm{s}$ 
\\ \hline
    $\rho_0$     & 1.204 & $\mathrm{kg}/\mathrm{m}^3$ \\ \hline
  \end{tabular}
  \caption{
    Measured values for the relevant material dependent constants for 
    dry air at temperature 20°C and pressure $p_0=1013\,\mathrm{hPa}$.
  }
\end{table}

The problem we are going to solve exactly is that of infinite cylinder
in the above described fluid with harmonic boundary conditions at
infinity. Since the cylinder itself is infinitely long we can simplify
the problem to a 2-dimensional one. This we assume throughout what
follows.

\section{Solution of the Stokes Equations for an Infinite Cylinder in
  a Viscous Harmonic Flow}

\cite{Stokes1950} has already calculated the effect of internal
friction of a harmonically moving fluid with angular frequency
$\omega$ on the motion of pendulums with radius $a$ in an
approximation that is valid for $a \sqrt{\omega/\nu_0} \ll 1$. In the
following we will give an analytical solution for the velocity field
that is valid for all $r>a$ and a range of frequencies $\omega$ with
Reynolds number $Re \ll 1$.

\begin{figure}[ht]
  \begin{center}
    \includegraphics[width=\columnwidth]{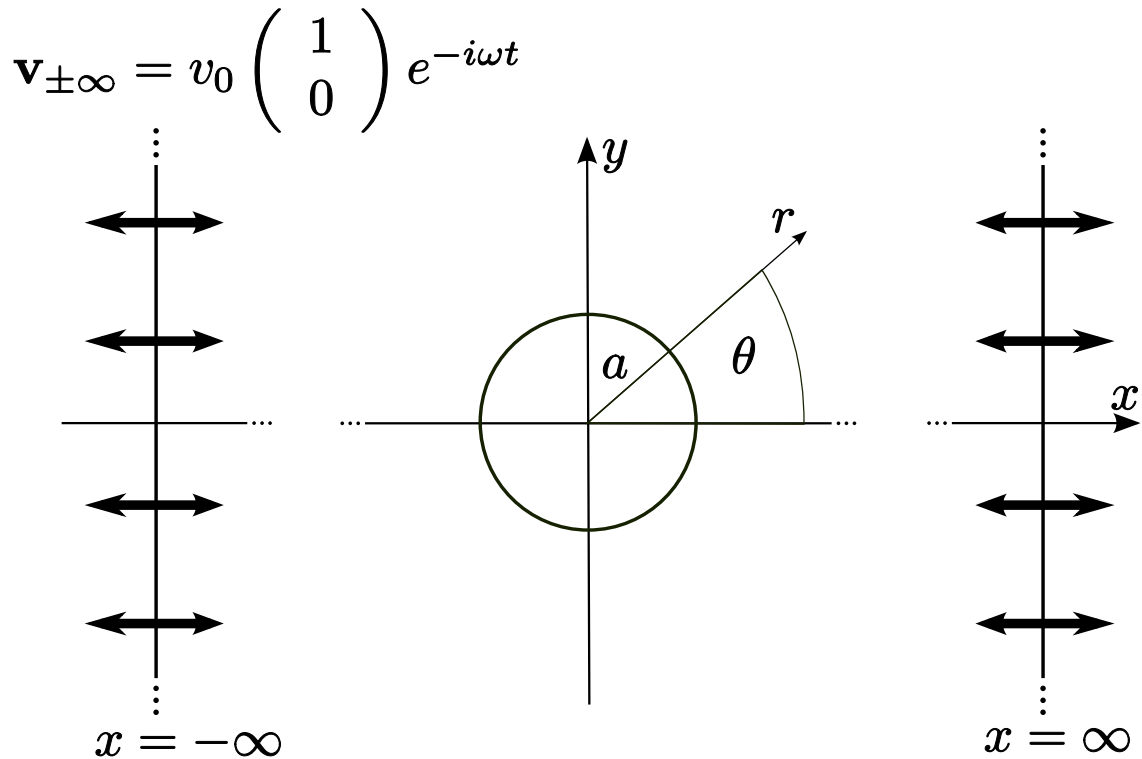}
    \caption{
      Cross-section through a cylinder with radius $a$ in a flow field
      that is harmonic at infinite distance from the cylinder. Angle
      $\theta$ and distance $r$ indicate the polar coordinates used
      here.  }
    \label{fig:ZeitabhStokes}
  \end{center}
\end{figure}

In practical work the stream velocity may well vary arbitrarily in
time but, to simplify the problem, we solve it for a \emph{harmonic}
flow field. Hence the stream velocity at infinite distance from the
cylinder is
\begin{flalign} \label{RBGeschwUnendlich}
  \vek{v}_{\pm \infty} = {v_{0}} \left ( \begin{array}{c} 1 \\ 0
    \end{array} \right) e^{-i \omega t} \,,
\end{flalign}
as shown in \figref{fig:ZeitabhStokes}.

For $Re \ll 1$ the non-linear term $\Kl \vek{v} \cdot \nabla \kl
\vek{v}$ is negligible and we can use the \emph{time-dependent Stokes
equation}
\begin{flalign} \label{ZeitabhStokes}
 \diffI{\vek{v}}{t} = - \frac{\gra p}{\rho_0} + \nu_{0} \lap \vek{v} 
\,.
\end{flalign}
It is \emph{linear} in $\vek{v}$. Constant volume forces such as
gravity do not play any role here since they only cause additional
gravitational pressure and do not change the form of the equations as
they can be incorporated into $\gra p$.

\subsection{Boundary Conditions}

The boundary condition \eqref{RBGeschwUnendlich} at infinite distance
from the cylinder is consistent with to the time-dependent Stokes
equation \eqref{ZeitabhStokes} as well as with the Navier-Stokes
equations \eqref{ImpulsInkomp}. This is important because at some
distance from the cylinder the time-dependent Stokes equation is not
valid any more.

On the cylinder surface we have no-slip boundary conditions,
\begin{flalign}
  \vek{v}\!\Kl a,t \kl =0 \,.
\end{flalign}
Using the boundary condition \eqref{RBGeschwUnendlich} at infinity we
can calculate the pressure at infinite distance from the origin. As
the velocity at infinity is homogeneous
\begin{flalign}
  \lap \vek{v}_{\infty} = 0
\end{flalign}
and $(\vek{v} \cdot \nabla) \vek{v}$
has been dropped we get, using \eqref{ImpulsInkomp},
\begin{flalign}
  \gra p_\infty = - \rho_0 \diffI{\vek{v}_\infty}{t} \,.
\end{flalign}
In view of \figref{fig:ZeitabhStokes} the boundary condition for the
pressure at infinity is therefore
\begin{flalign} \label{RBDruckUnendlich}
  p_{\infty} = i x \rho_0 \omega v_{0} e^{-i \omega t} \,.
\end{flalign}

\subsection{Solution of the Stokes Equation}

Applying the divergence to both sides of the time-dependent Stokes
equation \eqref{ZeitabhStokes} and taking advantage of
we get
\begin{flalign}
  \diffI{\div \vek{v}}{t} = - \frac{\div \gra p}{\rho_0} + \nu \lap
  \div \vek{v}
\end{flalign}
and as $\div \vek{v}=0$ we therefore find for the pressure
\begin{flalign} \label{StokesDruckGleichung}
 \lap p = 0 \,.
\end{flalign}
We first give a general solution to \eqref{StokesDruckGleichung} that
agrees with the boundary condition \eqref{RBDruckUnendlich} and the
symmetry of the problem.

In cylindrical coordinates, the boundary condition 
\eqref{RBDruckUnendlich}
for the pressure at infinity is
\begin{flalign} \label{RBDruckUnendZyl}
  p\!\Kl \infty, \theta \kl = i r \cos{\theta} \rho_0 \omega v_{0} 
e^{-i
    \omega t}
\end{flalign}
and \eqref{StokesDruckGleichung} becomes
\begin{flalign} \label{StokesDruckZyl} \lap p = \diffII{p}{r} +
  \frac{1}{r} \diffI{p}{r} + \frac{1}{r^2} \diffII{p}{\phi} +
  \diffII{p}{z} = 0 \,.
\end{flalign}
We give a general solution using the ansatz
\begin{flalign}
  p\!\Kl r,\theta,t \kl = R\!\Kl r \kl P\!\Kl \theta \kl T\!\Kl t \kl 
\,.
\end{flalign}
We get from \eqref{StokesDruckZyl}, as $R\!\Kl r \kl$ and $P\!\Kl
\theta \kl$ are functions depending on different variables, with a
constant $k$ independent of $r$ and $\theta$,
\begin{flalign} \label{AzimutalDruck}
  \diffII{P}{\theta} + k^2 P = 0
\end{flalign}
and
\begin{flalign} \label{RadialDruck}
  \frac{r^2}{R} \diffII{R}{r} + \frac{r}{R} \diffI{R}{r} - k^2 = 0 \,.
\end{flalign}
The general solution for $R\!\Kl r \kl$ of \eqref{RadialDruck} is
\begin{flalign} \label{RadialDruckLsg}
  R\!\Kl r \kl = \frac{A_{1,k}}{r^{k}} - A_{2,k} r^{k}
\end{flalign}
while for $P\!\Kl \theta \kl$
\begin{flalign} \label{AzimutalDruckLsg} 
  P\!\Kl \theta \kl = B_{1,k} \cos(k \theta) + B_{2,k} \sin( k \theta) \,.
\end{flalign} 
The pressure must be continuous and hence
\begin{flalign}
  P\!\Kl \theta \kl = P\!\Kl \theta + 2 \pi \kl
\end{flalign}
so that $k = 0, 1, 2, \dots$. As a general solution to
\eqref{StokesDruckZyl} we get
\begin{flalign}
  p\! \Kl r, \theta, t \kl = \sum_{k=0}^{\infty} \Ekl
  \frac{A_{1,k}}{r^{k}} - A_{2,k} r^{k} \ekl \Ekl B_{1,k} \cos{\Kl k
    \theta \kl}+ B_{2,k} \sin{\Kl k \theta \kl} \ekl T\!\Kl t \kl.  
\end{flalign}

To comply with \eqref{RBDruckUnendZyl} the solution cannot contain any
power of $r^k$ with $k > 1$ since the functions $\cos(k \theta)$ and
$\sin(k \theta)$ are linearly independent for all $k=0, 1, 2, \dots$.
As the Stokes equation is linear we set $T(t)=\exp(- i \omega t)$.
Expressions with $k=0$ are irrelevant since they only result in an
additive constant to the pressure. Possible solutions are therefore
\begin{flalign}
  p\!\Kl r,\theta,t \kl = \Gkl \sum_{k=1}^{\infty} \frac{1}{r^{k}}
  \Ekl C_{1,k} \cos{\Kl k \theta \kl} + C_{2,k} \sin{\Kl k \theta \kl}
  \ekl + i r \rho_{0} \omega v_{0} \cos{\theta} \gkl e^{-i \omega t} 
\,.
\end{flalign}
The problem is symmetric with respect to the $x$-axis and therefore
\begin{flalign}
  p\!\Kl r,\theta,t \kl &= p\!\Kl r, -\theta,t \kl
\end{flalign}
so that solutions depending on $\sin(k \theta)$ must be zero. As we
get another solution of the time-dependent Stokes equations by
applying the transformation $\vek{v} \to - \vek{v}$ and $p \to - p$,
which in our case is equivalent to mirroring with respect to the
$y$-axis, it holds that $p(-x,y)=-p(x,y)$ or, in polar coordinates,
\begin{flalign}
  p\!\Kl r,\theta,t \kl &= - p\!\Kl r, \pi-\theta,t \kl \,.
\end{flalign}
Since
\begin{flalign}
  \cos[ k \Kl \pi -\theta \kl ] = \Gkl
    \begin{array}{l}
       \ \ \, \cos( k \theta ) \quad \mbox{for $k$ even} \\
       -\cos( k \theta ) \quad \mbox{for $k$ odd,}
    \end{array} \right.
\end{flalign}
the solutions depending on $\cos(k \theta)$ with even $k$ must vanish.
The remaining most general solution that complies with the boundary
conditions is
\begin{flalign}
  p\!\Kl r,\theta,t \kl = \Ekl \sum_{k=1}^{\infty} \frac{1}{r^{k}}
  C_{k} \cos{\Kl k \theta \kl} + i r \rho_{0} \omega v_{0}
  \cos{\theta} \ekl e^{-i \omega t} \,.
\end{flalign}
We do not know yet the boundary condition for the pressure at the
surface of the cylinder as it results from the flow field. To simplify
further calculations we use the simple ansatz
\begin{flalign} \label{Druckhypothese} 
  p\!\Kl r,\theta,t \kl = \Ekl
  \frac{C}{r} + i r \rho_0 \omega v_{0} \ekl \cos{\theta} e^{-i \omega
    t}
\end{flalign}
and verify that it is consistent with the Stokes equations
\eqref{ZeitabhStokes}.

Since we are in two dimensions, the incompressibility condition reads in
polar coordinates
\begin{flalign} \label{KontiZyl}
  \div\vek{v} = \frac{1}{r} \diffI{\Kl r v_{r} \kl}{r} + \frac{1}{r}
  \diffI{v_{\theta}}{\theta} = 0
\end{flalign}
and the time-dependent Stokes equations are given by
\begin{flalign} \label{ZeitStokesZyl}
  \diffI{v_{r}}{t} &= - \frac{1}{\rho_0} \diffI{p}{r} + \nu \Ekl
  \diffII{v_{r}}{r} + \frac{1}{r^2} \diffII{v_{r}}{\theta} +
  \frac{1}{r} \diffI{v_{r}}{r} - \frac{2}{r^2}
  \diffI{v_{\theta}}{\theta} - \frac{v_{r}}{r^2} \ekl  \,,\\ 
  \diffI{v_{\theta}}{t} &= - \frac{1}{r \rho_0} \diffI{p}{\theta}  +
  \nu \Ekl \diffII{v_{\theta}}{r} + \frac{1}{r^2}
  \diffII{v_{\theta}}{\theta} + \frac{1}{r}
  \diffI{v_{\theta}}{r} + \frac{2}{r^2} \diffI{v_{r}}{\theta} -
  \frac{v_{\theta}}{r^2} \ekl \,.
\end{flalign}
We now substitute the pressure ansatz \eqref{Druckhypothese}. With the
help of \eqref{KontiZyl} the above equations decouple so as to give an
equation for $v_r$ alone,
\begin{flalign} \label{ZeitStokesVr}
   \diffI{v_{r}}{t} = \Ekl \frac{C}{\rho_{0} r^2}  - i \omega v_{0} 
\ekl \cos{\theta} e^{-i \omega t} + \nu \Ekl \diffII{v_{r}}{r} + 
\frac{1}{r^2} \diffII{v_{r}}{\theta} + \frac{3}{r} \diffI{v_{r}}{r} + 
\frac{v_{r}}{r^2} \ekl
\end{flalign}
while
\begin{flalign} \label{ZeitStokesVtheta}
  v_{\theta} = \int \diffI{\Kl r v_{r} \kl}{r}  d\theta \,.
\end{flalign}
A solution of the inhomogeneous equation \eqref{ZeitStokesVr} is
\begin{flalign} \label{vrinh}
  v_{r}\!\Kl r,\theta,t \kl =& \Ekl v_{0} + \frac{A J_{1}\!\Kl  j^{+}
    \beta r \kl}{\rho_{0} \omega r} + \frac{B K_{1}\!\Kl  j^{-} \beta
    r \kl}{\rho_{0} \omega r} + \frac{i C}{\rho_{0} \omega r^2}\ekl
  \cos{\theta} e^{-i \omega t}
\end{flalign}
with Bessel function $J_{1}$ of first order and second kind, modified
Bessel function $K_{1}$ of first order and second kind, and
\begin{flalign}
  j^{\pm} := \frac{1}{\sqrt{2}} \Kl 1 \pm i\kl\,,\ 
  \beta := \sqrt{\frac{\omega}{\nu_{0}}} \,.
\end{flalign}
The constants $A$, $B$ and $C$ are still to be determined.
The general solution of \eqref{ZeitStokesVr} is the sum of the general
solution of the homogeneous equation plus a special solution of the
inhomogeneous equation. If, however, the boundary conditions can be
satisfied by the inhomogeneous solution as given above we need not 
care about the homogeneous solution any more.

\begin{figure}[hbt]
 \begin{center}
   \includegraphics[width=\columnwidth]{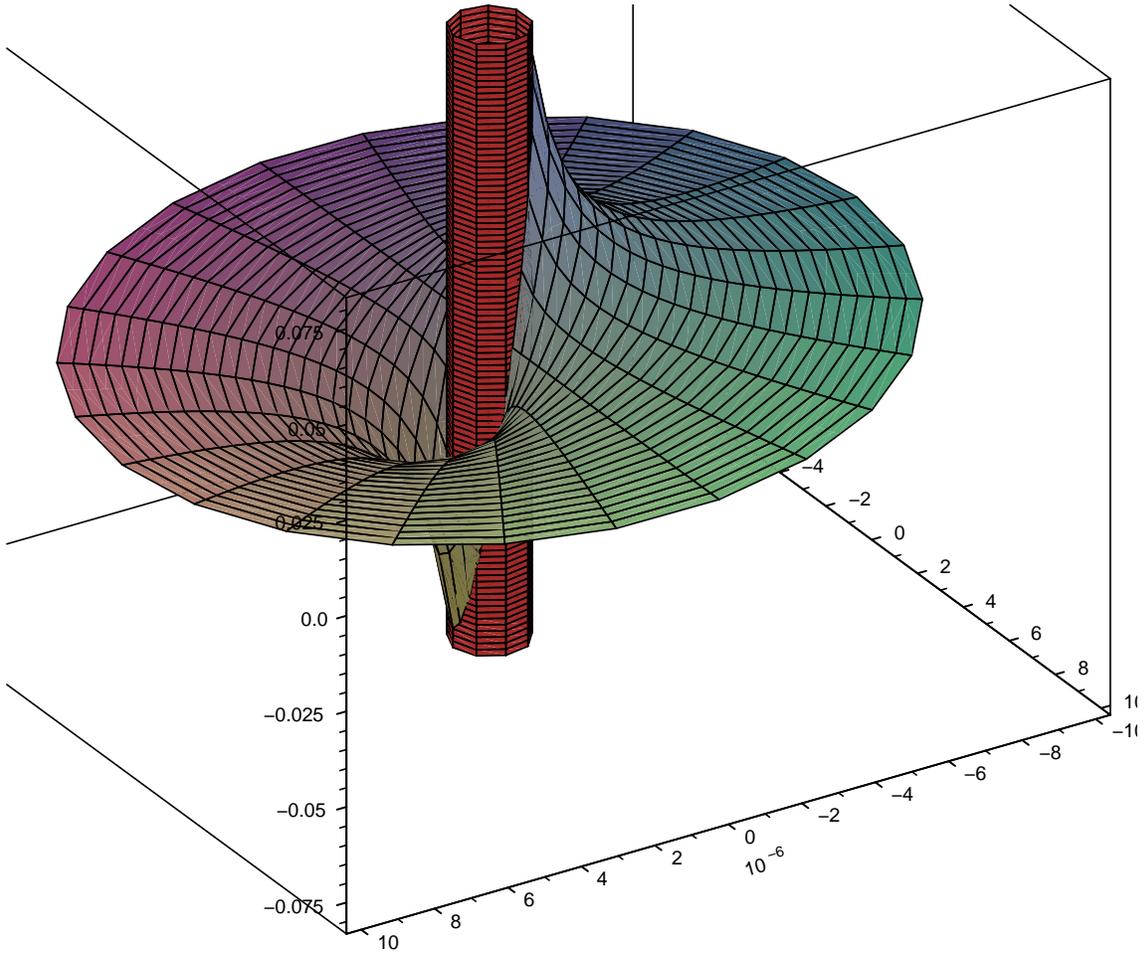}
 \caption{
   Pressure distribution $\Re[p(r,\theta,t=0)]$ around a cylinder
   (red) with radius $a=1\,\mu\mathrm{m}$ for frequency
   $f=\omega/2 \pi = 10\,\mathrm{Hz}$ in arbitrary units. The pressure
   decreases steeply within a few cylinder diameters. We assume dry
   air in all figures.
   \label{fig:p}
 }
 \end{center}
\end{figure}

\begin{figure}[hbt]
 \begin{center}
   \includegraphics[width=\columnwidth]{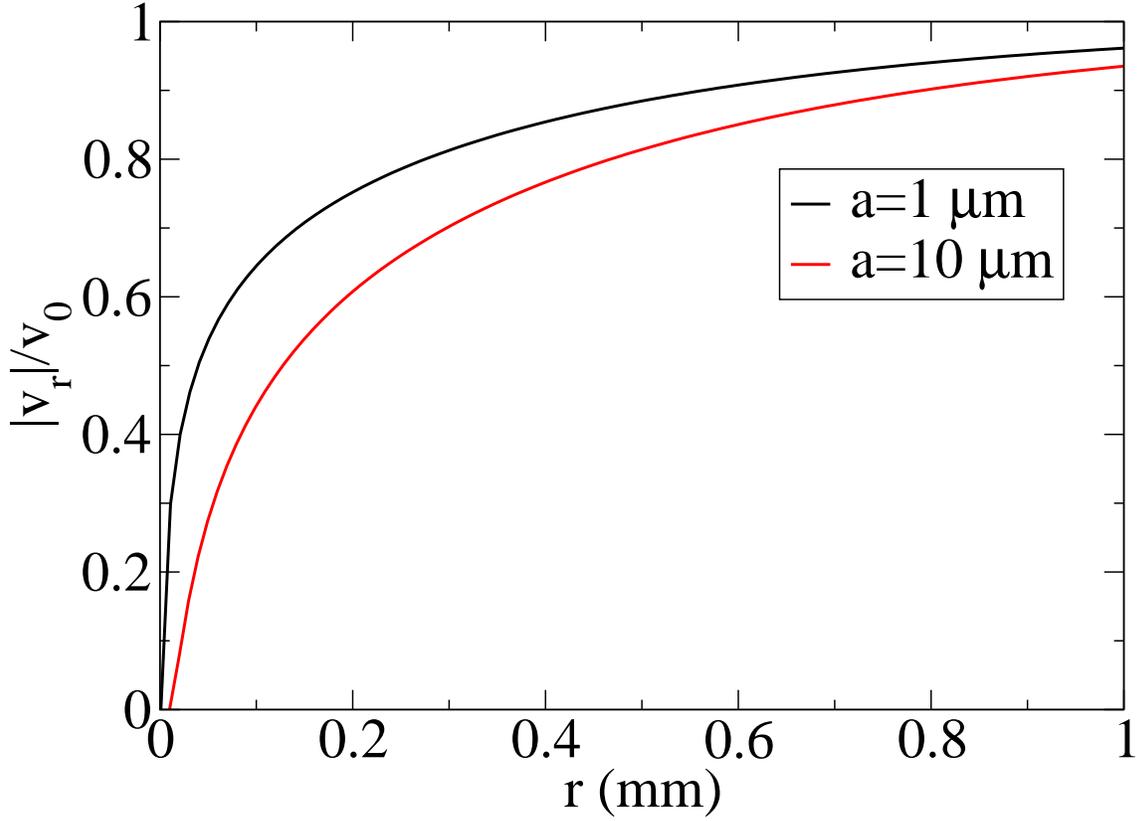}
 \caption{
   Amplitude of radial relative velocity $|v_r(r,\theta=0)|/v_0$
   of \eqref{vr}
   in dependence upon the distance $r$ to a cylinder with radius $a$ for
   frequency $f=\omega/2 \pi = 10\,\mathrm{Hz}$. The relative velocity
   approaches 1 for $r \to \infty$, as required by the boundary
   conditions. The thicker the cylinder, the farther the influence of the
   cylinder on the velocity field.
   \label{fig:vr_a}
 }
 \end{center}
\end{figure}

\begin{figure}[hbt]
 \begin{center}
   \includegraphics[width=\columnwidth]{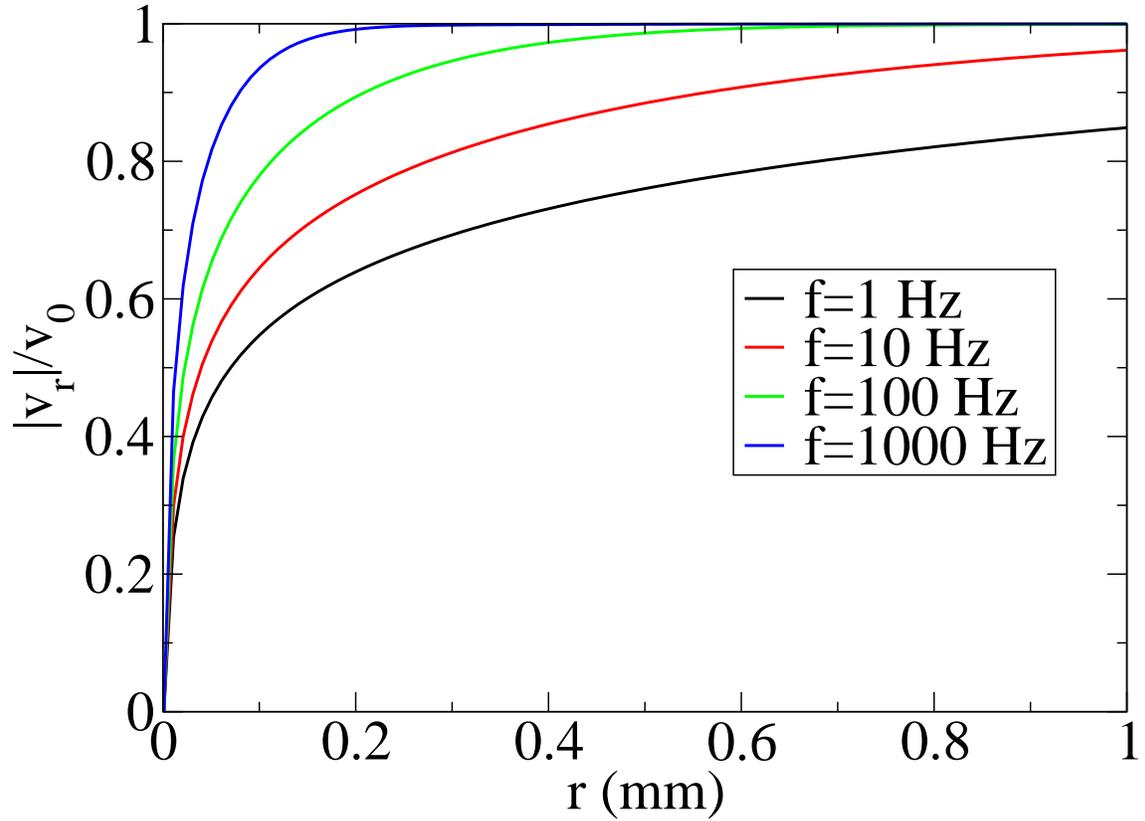}
 \caption{
   Amplitude of radial relative velocity $|v_r(r,\theta=0)|/v_0$
   of \eqref{vr} in dependence upon the distance $r$ to a cylinder with
   radius $a=1\,\mu\mathrm{m}$ for different frequencies. For high
   frequencies, the fluid velocity reaches its undisturbed value at
   much shorter distances than for lower frequencies.
   \label{fig:vr_f}
 }
 \end{center}
\end{figure}

\begin{figure}[hbt]
 \begin{center}
   \includegraphics[width=\columnwidth]{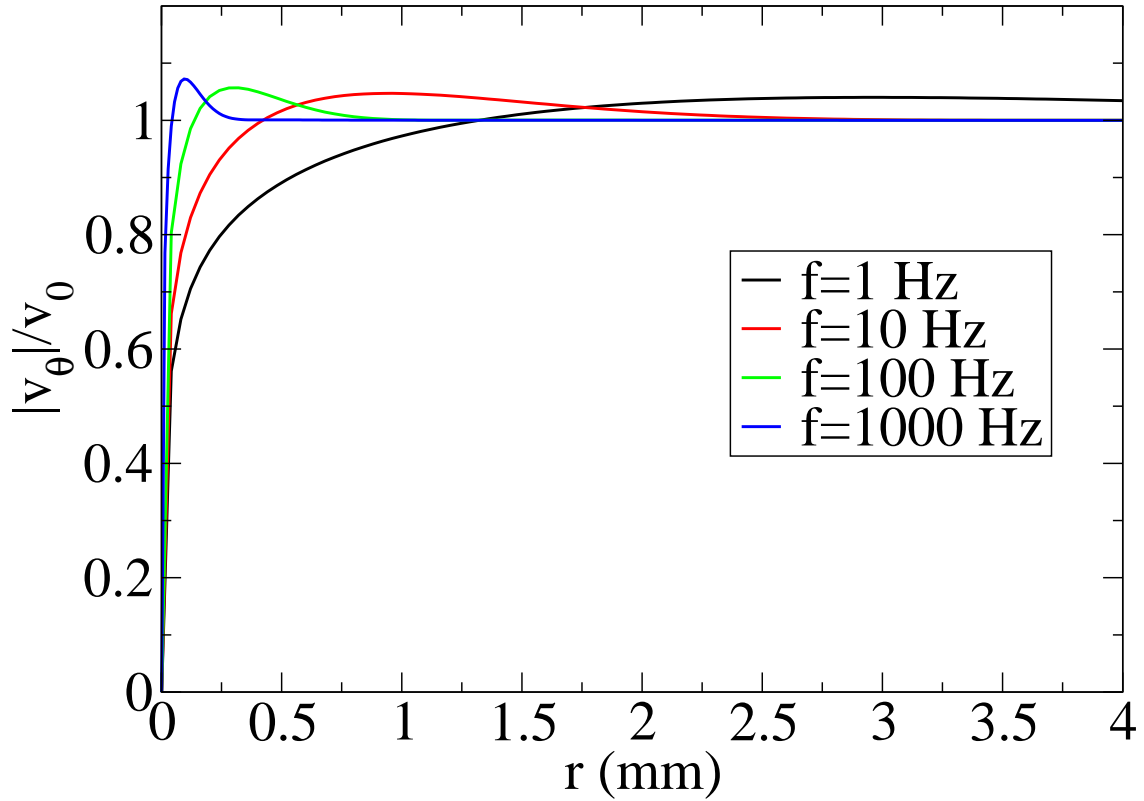}
 \caption{
   Amplitude of relative velocity $|v_\theta(r,\theta=\pi/2)|/v_0$
   of \eqref{vtheta} in dependence upon the distance $r$ to a cylinder with
   radius $a=1\,\mu\mathrm{m}$ for different frequencies. As the fluid
   must flow around the cylinder, relative velocities larger than 1 occur.
   For high frequencies, the fluid velocity reaches its undisturbed
   value at much shorter distances than for lower frequencies, as in
   \figref{fig:vr_f}.
   \label{fig:vtheta_f}
 }
 \end{center}
\end{figure}

The Bessel function of the second kind $J_{1}\!\Kl j^{+} \beta r \kl$
with complex argument
\begin{flalign}
  j^{+} \beta r = \frac{1}{\sqrt{2}} \Kl 1 + i\kl 
\sqrt{\frac{\omega}{\nu_{0}}} r
\end{flalign}
diverges to infinity as $r \rightarrow \infty$. Since the velocity has
to be finite at infinity we must set $A=0$. At the cylinder surface we
have the no-slip boundary condition
\begin{flalign} 
  v_{r}\!\Kl a,\theta,t \kl =& \Ekl v_{0} + \frac{B K_{1}\!\Kl  j^{-}
    \beta a \kl}{\rho_{0} \omega a} + \frac{i C}{\rho_{0} \omega
    a^2}\ekl \cos{\theta} e^{-i \omega t} = 0
\end{flalign}
and we therefore obtain
\begin{flalign}
  B = -\frac{\rho_{0} a^2 \omega v_{0} + i C}{a K_{1}\!\Kl  j^{-}
    \beta a \kl} \,.
\end{flalign}
Substituting the constant $B$ into \eqref{vrinh} we get
\begin{flalign}
  v_{r}\!\Kl r,\theta,t \kl =& \Gkl v_{0} + \frac{i C}{\rho_{0} \omega
    r^2} - \frac{a}{r} \Kl v_{0} + \frac{i C}{\rho_{0} \omega a^2} \kl
  \frac{K_{1}\!\Kl  j^{-} \beta r \kl}{K_{1}\!\Kl  j^{-} \beta a \kl}
  \gkl \cos{\theta} e^{-i \omega t}
\end{flalign}
while \eqref{ZeitStokesVtheta} and a little algebra provide us with
\begin{flalign}
  v_{\theta}\!\Kl r,\theta,t \kl = - \Ekl v_{0} - \frac{i C}{\rho_{0}
    \omega r^2} + \frac{a}{r} \Kl v_{0} +  \frac{i C}{\rho_{0} \omega
    a^2} \kl \frac{K_{1}\!\Kl j^{-} \beta r \kl}{K_{1}\!\Kl  j^{-}
    \beta a \kl} + \right. \\ \nonumber \left. + a \beta j^{-} \Kl
  v_{0} +  \frac{i C}{\rho_{0} \omega a^2} \kl \frac{K_{0}\!\Kl j^{-}
    \beta r \kl}{K_{1}\!\Kl  j^{-} \beta a \kl} \ekl \sin{\theta}
  e^{-i \omega t} \,.
\end{flalign}
There is one last constant $C$ that still has to be determined. We solve
the no-slip boundary condition $v_{\theta}\!\Kl a,\theta,t\kl =0$ at
the cylinder surface for $C$ and find
\begin{flalign}
  C = i \rho_{0} a^2 \omega v_{0} \Kl 1 + \frac{2 j^{+}}{\beta a}
  \frac{K_{1}\!\Kl j^{-} \beta a \kl}{K_{0}\!\Kl j^{-} \beta a \kl}
  \kl \,. 
\end{flalign}
As an abbreviation we introduce
\begin{flalign}
  f(r) := \frac{2 j^{+} K_{1}\!\Kl j^{-} \beta r \kl}{
    K_{0}\!\Kl j^{-} \beta a \kl}
\end{flalign}
and with
\begin{flalign}
  \beta = \sqrt{\frac{\omega}{\nu_{0}}}
\end{flalign}
we finally obtain the full solution for $r > a$,
\begin{flalign}
  v_{r}\!\Kl r,\theta,t \kl &= v_{0} \cos(\theta) e^{-i \omega t} \Ekl
  1-\frac{a^2}{r^2} + \frac{f(r)}{\beta r} - \frac{a f(a)}{\beta r^2}
  \ekl \,, \label{vr} \\
  v_{\theta}\!\Kl r,\theta,t \kl &= v_0 \sin(\theta) e^{-i \omega t}
  \Ekl - 1 - \frac{a^2}{r^2} + 2 \frac{K_{0}\!\Kl j^{-} \beta r
    \kl}{K_{0}\!\Kl j^{-} \beta a
    \kl} + \frac{f(r)}{\beta r} - \frac{a f(a)}{\beta r^2} \ekl
  \label{vtheta} \,, \\
  p(r,\theta,t) &= i v_0 \rho_0 \omega \cos(\theta) e^{-i \omega t}
  \Ekl r + \frac{a^2}{r} + \frac{a f(a)}{\beta r} \ekl \label{p} \,.
\end{flalign}
\Figref{fig:p} shows the pressure distribution around the cylinder.
Figures \ref{fig:vr_a}, \ref{fig:vr_f}, and \ref{fig:vtheta_f} contain
plots of the fluid velocity. The fluid velocity always reaches its
undisturbed value $v_0$ for $r \to \infty$, as required by the
boundary conditions at infinity. For high frequencies, the fluid 
velocity reaches
its undisturbed value for much shorter distances than for lower
frequencies. 

\clearpage

\section{Force on a Cylinder}

\begin{figure}[hbt]
 \begin{center}
   \includegraphics[width=\columnwidth]{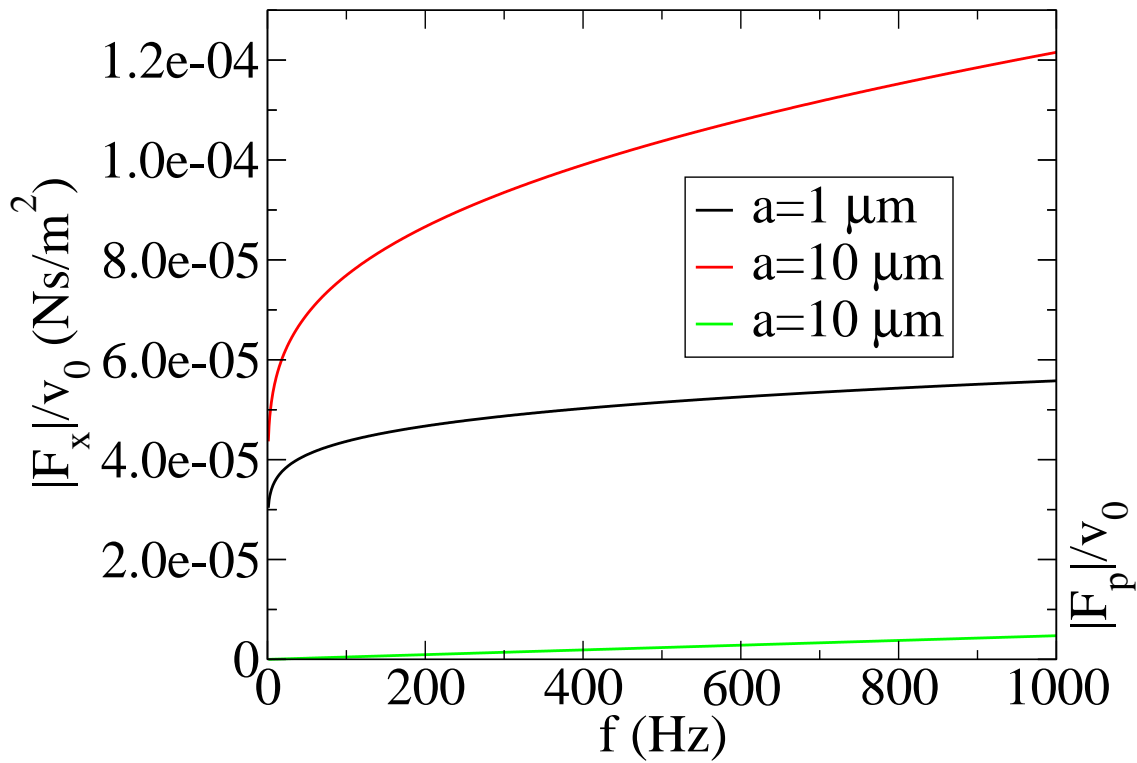}
 \caption{
   The amplitude of the force $|F_x|/v_0$ per cylinder length relative
   to fluid velocity $v_0$ acting upon a cylinder that stands still;
   cf.  \eqref{F}. The force is plotted for different cylinder radii
   $a$ in dependence upon the frequency $f$. For a thin cylinder
   ($a=1\,\mu\mathrm{m}$) the force is more or less independent of the
   frequency in the range 100--1000\,Hz. For thicker cylinders
   ($a=10\,\mu\mathrm{m}$) the force increases with frequency. The
   ``buoyancy'' force $F_p$ (green graph) of \eqref{Fp} is negligible
   in comparison to viscous forces.
   \label{fig:F}
 }
 \end{center}
\end{figure}

The force $\vek{f}$ acting on a surface element is given by stress
tensor $\vek{\Pi}$ operating on the normal vector $\vek{n}$ of the
surface,
\begin{flalign}
 \vek{f} =  \vek{\Pi} \vek{n} \,.
 \label{fpin}
\end{flalign} 
In polar coordinates the stress tensor is given by \citep[Art. 328a]{Lamb1932}
\begin{flalign}
  \vek{\Pi} =  \left( \begin{array}{cc} 
	-p + 2 \mu_{0} \partial_r v_r &
        \mu_{0} \left( \frac{1}{r} \partial_\theta v_r + \partial_r
          v_\theta - \frac{1}{r} v_\theta \right)  \\
        \mu_{0} \left( \frac{1}{r} \partial_\theta v_r + \partial_r
          v_\theta - \frac{1}{r} v_\theta \right)  & -p + 2 \mu_{0}
        \left(\frac{1}{r} \partial_\theta v_\theta + \frac{1}{r} v_r
        \right)  \\
    \end{array} \right) \,.
  \label{pi}
\end{flalign}
The normal vector on the cylinder surface reduced to a circle is
\begin{flalign}
  \vek{n} = \Kl \begin{array}{c} \cos \theta \\ \sin \theta 
\end{array} \kl \,.
\end{flalign}
Because of \eqref{fpin} and \eqref{pi}, the force $f_x$ per area in
$x$-direction is given by
\begin{flalign}
  f_x = \Kl-p + 2 \mu_{0} \partial_r v_r \kl \cos{\theta} - \mu_{0}
  \Kl \frac{1}{r} \partial_\theta v_r + \partial_r v_\theta -
  \frac{1}{r} v_\theta \kl \sin{\theta} \,.
\end{flalign}
The total force $F_y$ in $y$-direction is zero because of symmetry.
The total force $F_x$ in $x$-direction per cylinder length is
\begin{flalign}
  F_x = \int_{0}^{2 \pi} \left[ 
	\Kl-p + 2 \mu_{0} \partial_r v_r \kl \cos{\theta} - \mu_{0}
        \Kl \frac{1}{r} \partial_\theta v_r + \partial_r v_\theta -
        \frac{1}{r} v_\theta \kl \sin{\theta} \right]
    a \, d\theta \,. \nonumber
\end{flalign}
Substituting of the solution (\ref{vr}--\ref{p}) and integration gives
\begin{flalign}
  F_x = - i 2 \pi \rho_{0} v_{0} \omega a \Ekl a + \frac{f(a)}{\beta}
  \ekl \ e^{-i \omega t} \,.
  \label{F}
\end{flalign}
\Figref{fig:F} shows the force acting on a cylinder in dependence upon
the fluid frequency. For low enough frequencies and thin cylinders
forces that are independent of viscosity, so-called ``buoyancy''
forces,
\begin{flalign}
  F_p = - i 2 \pi \rho_{0} v_{0} \omega a^2 \ e^{-i \omega t} \,,
  \label{Fp}
\end{flalign}
can be neglected as compared to viscous forces. When calculating the
force on an oscillating cylinder with angular frequency $\omega$
additional buoyancy forces proportional to $\rho_{0} v_{0} \omega a^2$
occur \citep{Panton2005a}. If these are negligible, however, we need
only consider the \emph{relative} velocity $v_0$ between cylinder and
fluid even in case of an oscillating cylinder so as to find
\begin{flalign}
  F_x \approx - i 2 \pi \rho_{0} \, v_0 \, \omega \,
  \frac{a f(a)}{\beta} \ e^{-i \omega t} \,.
\end{flalign}

\clearpage

\section{Discussion}

For a very long cylinder in the low Reynolds number regime
time-dependent incompressible Stokes equations hold. For harmonic
flow, these can be solved analytically giving the flow and pressure
field around a cylinder.  However, the approximation of an infinitely
long cylinder is only fulfilled when the stream velocity converges to
its undisturbed constant value within a short range, so that the
cylinder must be much longer than the distance where the velocity has
reached about 90\,\% of the undisturbed value. This precondition is
not fulfilled for very low frequencies as shown in
Figs.~\ref{fig:vr_f} and \ref{fig:vtheta_f}.

The velocity field and viscous forces can be calculated analytically
using the linear time-dependent Stokes equations as an approximation.
In contrast to Stokes (1851), who computed the velocity field in the
direct neighborhood of a cylinder, and, hence, the force for small
radius $a$, we provide the \emph{full velocity field} for \emph{all}
$r > a$.

 

\end{document}